\documentclass[12pt]{aipproc} 
\layoutstyle{6x9}

\newcommand\doingARLO[2][]{\ifx\mmref\undefined #1\else #2\fi}

\begin{document}

\begin{flushright}
CLNS~01/1760\\
{\tt hep-ph/0110093}
\end{flushright}
\vspace{-0.9truecm}

\title{Aspects of QCD Factorization\footnote{Talk presented at the 
{\em Int.\ Workshop on QCD: Theory and Experiment\/} (Martina Franca, 
Italy, 16--20 June 2001), and at the {\em 9$^{th}$ Int.\ Symposium on 
Heavy Flavour Physics\/} (Pasadena, CA, 10--13 September 2001)}}

\author{Matthias Neubert}{
address={Newman Laboratory of Nuclear Studies, Cornell University, 
Ithaca, NY 14853, USA},
email={neubert@mail.lns.cornell.edu}}

\begin{abstract}
The QCD factorization approach provides the theoretical basis for a
systematic analysis of nonleptonic decay amplitudes of $B$ mesons in
the heavy-quark limit. After recalling the basic ideas underlying this
formalism, several tests of QCD factorization in the decays 
$B\to D^{(*)} L$, $B\to K^*\gamma$, and $B\to\pi K,\pi\pi$ are 
discussed. It is then illustrated how factorization can be used to 
obtain new constraints on the parameters of the unitarity triangle.
\end{abstract}

\maketitle

\section{Introduction}

In many years of intense experimental and theoretical investigations
the flavor sector of the Standard Model has been explored in great
detail by studying mixing and weak decays of $B$ mesons and kaons. CP
violation has been observed in $K$--$\bar K$ mixing (1964), $K\to\pi\pi$ 
decays (1999), and most recently in the interference of mixing and decay 
in $B\to J/\psi\,K$ (2001). There is now compelling evidence that the 
Cabibbo--Kobayashi--Maskawa (CKM) mechanism accounts for the dominant 
source of CP violation in low-energy hadronic weak interactions. Most 
notably, the discovery of a large CP asymmetry in the $B$ system has 
established that CP is not an approximate symmetry of Nature. Rather, 
the smallness of CP-violating effects in kaon (and charm) physics 
reflects the hierarchy of CKM matrix elements. 

Measurements of $|V_{cb}|$ and $|V_{ub}|$ in semileptonic $B$ decays 
and of the magnitude and phase of $V_{td}$ in $K$--$\bar K$ mixing, 
$B_{d,s}$--$\bar B_{d,s}$ mixing, and $B\to J/\psi\,K$ decays has helped
to determine the parameters of the unitarity triangle
$V_{ub}^* V_{ud}+V_{cb}^* V_{cd}+V_{tb}^* V_{td}=0$ with good accuracy. 
The current values obtained at 95\% confidence level are 
$\bar\rho=0.21\pm 0.12$, $\bar\eta=0.38\pm 0.11$ for the coordinates of 
the apex of the (rescaled) triangle, and $\sin 2\beta=0.74\pm 0.15$, 
$\sin2\alpha=-0.14\pm 0.57$, $\gamma=(61\pm 16)^\circ$ for its angles 
\cite{Hocker:2001xe}. These studies have established the existence of a 
CP-violating phase in the top sector of the CKM matrix, i.e., 
$\mbox{Im}(V_{td}^2)\ne 0$. The next step in testing the CKM paradigm 
must be to explore the CP-violating phase in the bottom sector, i.e., 
$\gamma=\mbox{arg}(V_{ub}^*)\ne 0$. In the Standard Model the two 
phases are, of course, related to each other. However, there is still 
plenty of room for New Physics to affect the magnitude of flavor 
violations in both mixing and weak decays (see, e.g., 
\cite{Kagan:2000wm}). In particular, the present upper bound on $\gamma$ 
is derived from the experimental limit on $B_s$--$\bar B_s$ mixing, 
which has not yet been seen experimentally and could well be affected 
by New Physics. 

\begin{figure}
\label{fig:treepeng}
\includegraphics[height=.15\textheight]{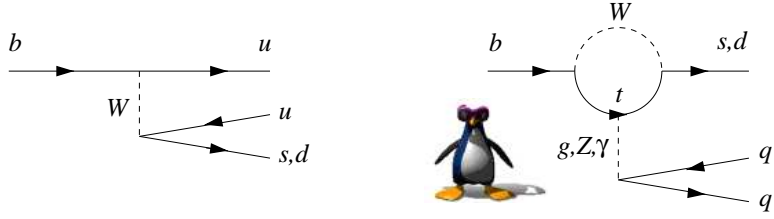}
\caption{Tree and penguin topologies in charmless hadronic $B$ decays.}
\end{figure}

Common lore says that measurements of $\gamma$ are difficult. Several
``theoretically clean''\footnote{In this area of flavor physics many 
practitioners would consider a method to be ``theoretically clean'' only
if it exclusively relies on elementary geometry (amplitude triangles) 
and, perhaps, isospin symmetry. We adopt the rationale followed in most 
other branches of high-energy physics and call a method theoretically 
clean if it relies on systematic expansions in small parameters. The 
methods discussed later in this talk are theoretically clean in this 
wider sense.}
determinations of this phase have been suggested (see, e.g., 
\cite{Atwood:1997ci,Dunietz:1988bv}), which are extremely challenging 
experimentally. Likewise, ``clean'' measurements of 
$\alpha=\pi-\beta-\gamma$ \cite{Gronau:1990ka,Snyder:1993mx} are very
difficult. It is more accessible experimentally to probe $\gamma$ (and 
$\alpha$) via the sizeable tree--penguin interference in charmless 
hadronic decays such as $B\to\pi K$ and $B\to\pi\pi$. The basic decay 
topologies contributing to these modes are shown in 
Figure~\ref{fig:treepeng}. Experiment shows that the tree-to-penguin 
ratios in the two cases are roughly $|T/P|_{\pi K}\approx 0.2$ and 
$|P/T|_{\pi\pi}\approx 0.3$, indicating a sizeable amplitude 
interference. It is important that the relative weak phase between the 
two amplitudes can be probed not only via CP asymmetry measurements 
($\sim\sin\gamma$), but also via measurements of CP-averaged branching 
fractions ($\sim\cos\gamma$). Extracting information about CKM 
parameters from the analysis of nonleptonic $B$ decays is a challenge to
theory, since it requires some level of control over hadronic physics, 
including strong-interaction phases. Such challenges, combined with the 
importance of the issue, is what triggers theoretical progress.

\section{QCD Factorization}

Hadronic weak decay amplitudes simplify greatly in the heavy-quark limit
$m_b\gg\Lambda_{\rm QCD}$. This statement should not surprise those who 
have followed the dramatic advances in our theoretical understanding of
$B$ physics in the past decade. Many areas of $B$ physics, from 
spectroscopy to exclusive semileptonic decays to inclusive rates and 
lifetimes, can now be systematically analyzed using heavy-quark 
expansions. Yet, the more complicated exclusive nonleptonic decays have
long resisted any theoretical progress. The technical reason is that, 
whereas in most other applications of heavy-quark expansions one 
proceeds by integrating out heavy fields (leading to local operator 
product expansions), in the case of nonleptonic decays the large scale 
$m_b$ enters as the energy carried by light fields. Therefore, in 
addition to hard and soft subprocesses collinear degrees of freedom 
become important. This complicates the understanding of hadronic decay 
amplitudes using the language of effective field theory. (Yet, very 
significant progress towards an effective field-theory description of 
nonleptonic decays has been made recently with the establishment of a 
``collinear--soft effective theory'' \cite{Bauer:2001yr}. The reader is 
referred to these papers for more details on this important development.) 

The importance of the heavy-quark limit is based on the physical idea
of color transparency \cite{Bjorken:1989kk,Dugan:1991de,Politzer:1991au}.
A fast-moving light meson (such as a pion) produced in a point-like
source (a local operator in the effective weak Hamiltonian) decouples 
from soft QCD interactions. More precisely, the couplings of soft gluons 
to such a system can be analyzed using a multipole expansion, and the 
first contribution (from the color dipole) is suppressed by a power of 
$\Lambda_{\rm QCD}/m_b$. The QCD factorization approach provides a 
systematic, model-independent implementation of this idea
\cite{Beneke:1999br,Beneke:2000ry}. It gives rigorous results in the 
heavy-quark limit, which are valid to leading power in 
$\Lambda_{\rm QCD}/m_b$ but to all orders of perturbation theory. 
Having obtained control over nonleptonic decays in the heavy-quark limit 
is a tremendous advance. We are now able to talk about power corrections
to a well-defined and calculable limiting case, which captures a 
substantial part of the physics in these complicated processes.

\begin{figure}
\label{fig:fact}
\includegraphics[width=.43\textwidth]{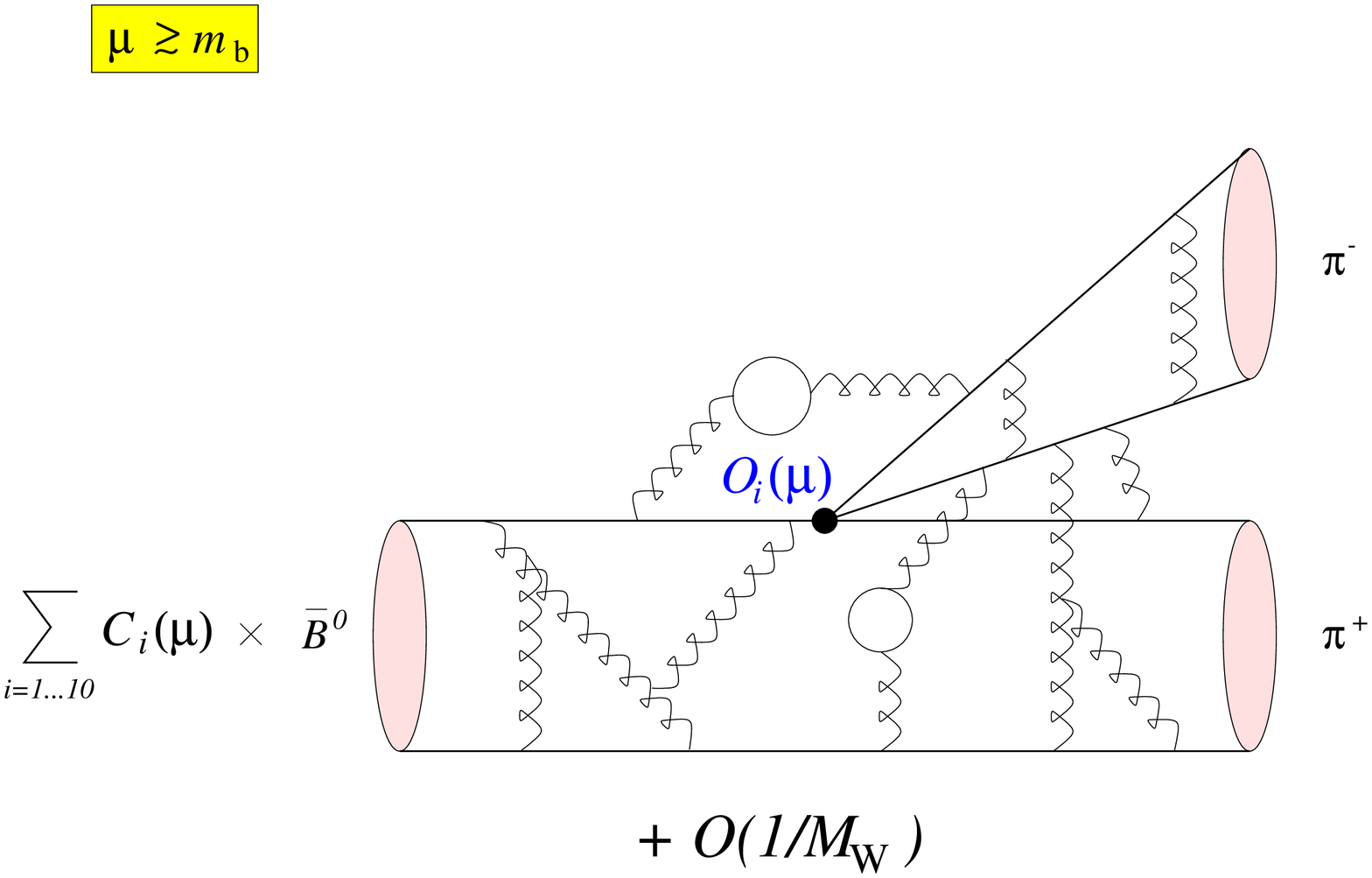}
\hspace{0.2truecm}
\includegraphics[width=.53\textwidth]{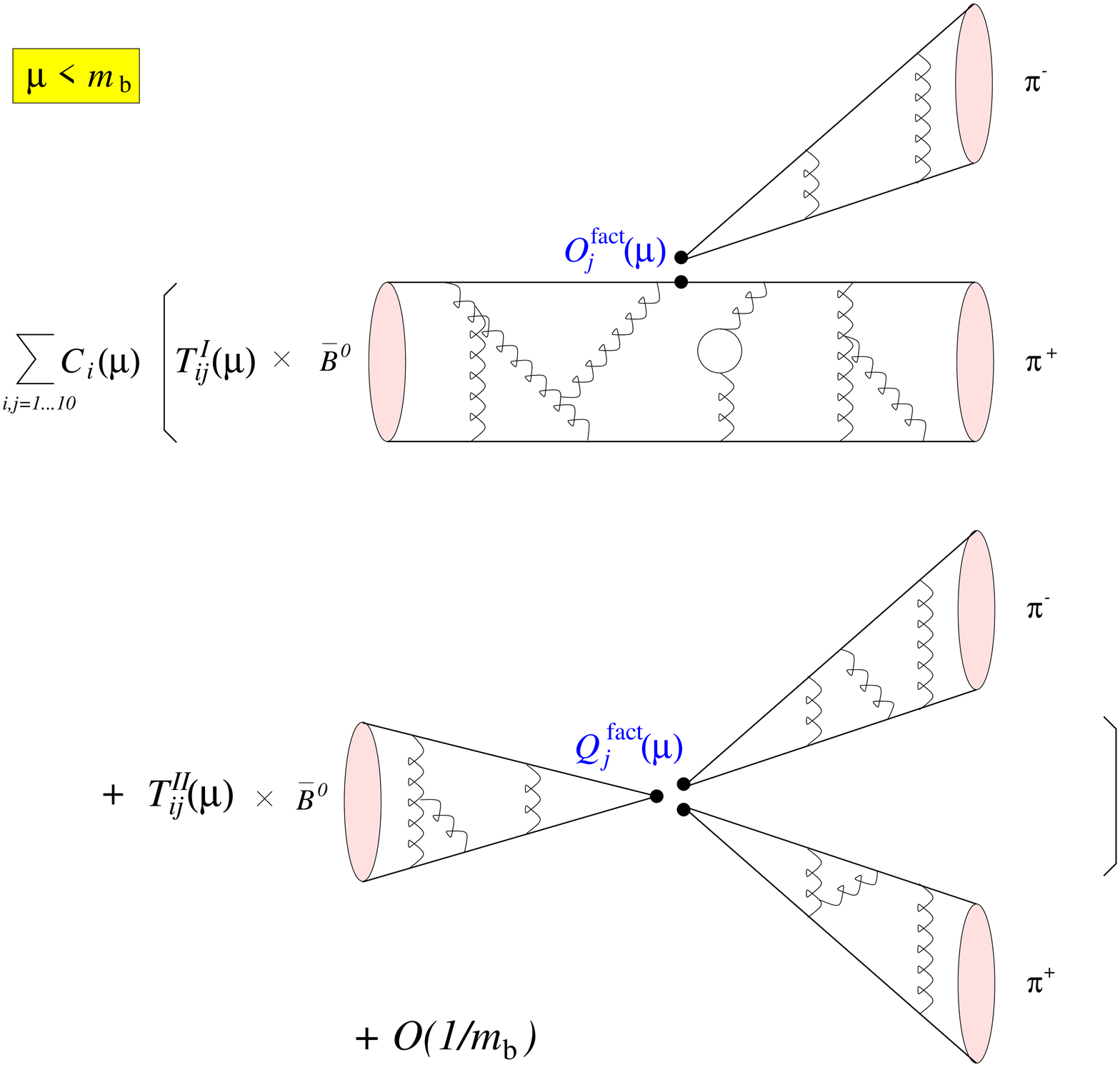}
\caption{Factorization of short- and long-distance contributions in 
hadronic $B$ decays. Left: Factorization of short-distance effects into
Wilson coefficients of the effective weak Hamiltonian. Right: 
Factorization of hard ``nonfactorizable'' gluon exchanges into 
hard-scattering kernels (QCD factorization).}
\end{figure}

The workings of QCD factorization can best be illustrated with the 
cartoons shown in Figure~\ref{fig:fact}. The first graph shows the
well-known concept of an effective weak Hamiltonian obtained by 
integrating out the heavy fields of the top quark and weak gauge bosons 
from the Standard Model Lagrangian. This introduces new effective 
interactions mediated by local operators $O_i(\mu)$ (typically four-quark 
operators) multiplied by calculable running coupling constants $C_i(\mu)$
called Wilson coefficients. This reduction in complexity (nonlocal heavy 
particle exchanges $\to$ local effective interactions) is exact up to 
corrections suppressed by inverse powers of the heavy mass scales. The 
resulting picture at scales at or above $m_b$ is, however, still rather 
complicated, since gluon exchange is possible between any of the quarks 
in the external meson states. Additional simplifications occur when the 
renormalization scale $\mu$ is lowered below the scale $m_b$. Then color 
transparency comes to play and implies systematic cancellations of soft 
and collinear gluon exchanges. As a result, all ``nonfactorizable'' 
exchanges, i.e., gluons connecting the light meson at the ``upper'' 
vertex to the remaining mesons, are dominated by virtualities of order 
$m_b$ and can be calculated. Their effects are absorbed into a new set 
of running couplings $T_{ij}^{I,II}(\mu)$ called hard-scattering 
kernels, as shown in the two graphs on the right-hand side. What remains 
are ``factorized'' four-quark and six-quark operators 
$O_j^{\rm fact}(\mu)$ and $Q_j^{\rm fact}(\mu)$, whose matrix elements 
can be expressed in terms of form factors, decay constants and 
light-cone distribution amplitudes. As before, the reduction in 
complexity (local four-quark operators $\to$ ``factorized'' operators) 
is exact up to corrections suppressed by inverse powers of the heavy 
scale, now set by the mass of the $b$ quark.

The factorization formula is valid in all cases where the meson at the
``upper'' vertex is light, meaning that its mass is much smaller than 
the $b$-quark mass. The second term in the factorization formula (the 
term involving ``factorized'' six-quark operators) gives a 
power-suppressed contribution when the final-state meson at the 
``lower'' vertex is a heavy meson (i.e., a charm meson), but its 
contribution is of leading power if this meson is also light. Aspects of 
this power counting will be discussed in more detail later.

Factorization is a property of decay amplitudes in the heavy-quark limit. 
Comparing the magnitude of ``nonfactorizable'' effects in kaon, charm
and beauty decays, there can be little doubt about the relevance of the 
heavy-quark limit to understanding nonleptonic processes 
\cite{Neubert:2001sj}. Yet, for phenomenological applications it is 
important to explore the structure of at least the leading 
power-suppressed corrections. While no complete classification of such 
corrections has been given to date, several classes of power-suppressed 
terms have been analyzed and their effects estimated. These estimates 
(with conservative errors) have been implemented in the phenomenological 
applications to be discussed later in this talk. Specifically, the 
corrections that have been analyzed are ``chirally-enhanced'' power 
corrections \cite{Beneke:1999br}, weak annihilation contributions 
\cite{Beneke:2000ry,Beneke:2001ev}, and power corrections due to 
nonfactorizable soft gluon exchange 
\cite{Khodjamirian:2001mi,Burrell:2001pf,Becher:2001hu}. With the 
exception of the ``chirally-enhanced'' terms, no unusually large power 
corrections (i.e., corrections exceeding the naive expectation of 
5--10\%) have been identified so far. Nevertheless, it is important to 
refine and extend the estimates of power corrections. Fortunately, the 
QCD factorization approach has a wide range of applicability and makes 
many testable predictions. Ultimately, therefore, the data will give us
conclusive evidence on the relevance of power-suppressed effects. Many 
tests can, in fact, already be done using existing data. Several 
examples will now be discussed in detail.

\subsection{Tests of Factorization in $B\to D^{(*)} L\,$ Decays} 

In $B$ decays into a heavy--light final state, when the light meson is 
produced at the ``upper'' vertex, the factorization formula assumes its
simplest form. Then only the form factor term (the first graph on the 
right-hand side in Figure~\ref{fig:fact}) contributes at leading power. 
This is also the place where QCD factorization is best established 
theoretically. In \cite{Beneke:2000ry}, the systematic cancellation of 
soft and collinear singularities was demonstrated explicitly at two-loop 
order. The proof of these cancellations has recently been extended to 
all orders in perturbation theory \cite{Bauer:2001cu}. In order to 
complete a rigorous proof of factorization one would still have to show 
that the hard-scattering kernels are free of endpoint singularities 
stronger than $1/x$ or $1/(1-x)$ as one of the quarks in the light 
meson becomes a soft parton. It has been demonstrated that the kernels 
tend to a constant (modulo logarithms) at the endpoints in the so-called 
``large-$\beta_0$ limit'' of QCD, i.e., to order 
$\beta_0^{n-1}\alpha_s^n$ for arbitrary $n$ in perturbation theory 
\cite{Becher:2001hu}. However, it is an open question whether such a 
smooth behavior persists in higher orders of full QCD. 

Let us first consider the decays $\bar B^0\to D^{(*)+} L^-$, where $L$ 
denotes a light meson. In this case the flavor content of the final 
state is such that the light meson can only be produced at the ``upper'' 
vertex, so factorization applies. One finds that process-dependent 
``nonfactorizable'' corrections from hard gluon exchange, though 
present, are numerically very small. All nontrivial QCD effects in the 
decay amplitudes are then described by a quasi-universal coefficient 
$|a_1(D^{(*)}L)|=1.05\pm 0.02+O(\Lambda_{\rm QCD}/m_b)$ 
\cite{Beneke:2000ry}. For a given decay channel this coefficient can be 
determined experimentally from the ratio \cite{Bjorken:1989kk}
\[
   \frac{\Gamma(\bar B^0\to D^{*+}L^-)}
        {d\Gamma(\bar B^0\to D^{*+}l^-\nu)/dq^2 \big|_{q^2=m_L^2}}
   = 6\pi^2 |V_{ud}|^2 f_L^2\,|a_1(D^{(*)}L)|^2 \,.
\]
Using CLEO data one obtains $|a_1(D^*\pi)|=1.08\pm 0.07$, 
$|a_1(D^*\rho)|=1.09\pm 0.10$, and $|a_1(D^* a_1)|=1.08\pm 0.11$, in 
good agreement with theory. This is a first indication that power 
corrections in these modes are under control, but more precise data are 
required for a firm conclusion. Other tests of factorization in $B$ 
decays to heavy--light final states have been discussed in 
\cite{Beneke:2000ry,Ligeti:2001dk,Diehl:2001xe}.

Recently, the experimental observation of unexpectedly large rates for 
color-suppressed decays such as $\bar B^0\to D^{0(*)}\pi^0$ 
\cite{CLEO,Abe:2001pd} has attracted some attention. QCD factorization 
does not allow us to calculate the amplitudes for these processes in a 
reliable way. It predicts that these amplitudes are power-suppressed 
with respect to the corresponding $\bar B^0\to D^{+(*)}\pi^-$ amplitudes, 
but only by one power of $\Lambda_{\rm QCD}/m_c$. Specifically, the 
prediction is that a certain ratio of isospin amplitudes approaches 
unity in the heavy-quark limit: $A_{1/2}/(\sqrt2 A_{3/2})
=1+O(\Lambda_{\rm QCD}/m_c)$ \cite{Beneke:2000ry}. Considering that 
charm is not a particularly heavy quark, we find that this scaling law 
is respected by the experimental data, which give
$A_{1/2}/(\sqrt2 A_{3/2})=(0.70\pm 0.11)\,e^{\pm i(27\pm 7)^\circ}$ for
$B\to D\,\pi$ and $(0.72\pm 0.08)\,e^{\pm i(21\pm 8)^\circ}$ for
$B\to D^*\pi$ \cite{Neubert:2001sj}. Assuming the hierarchy 
$m_b\gg m_c>\Lambda_{\rm QCD}$, a rough theoretical estimate of the 
amplitude ratio, $A_{1/2}/(\sqrt2 A_{3/2})\sim 0.75\,e^{-15^\circ\,i}$,
had been obtained prior to the observation of the color-suppressed decays 
\cite{Beneke:2000ry}. It anticipated the correct order of magnitude of 
the deviation from the heavy-quark limit. 

\subsection{Tests of Factorization in $B\to K^*\gamma\,$ Decays} 

The QCD factorization approach not only applies to nonleptonic decays, 
but also to other exclusive processes such as $B\to V\gamma$ and 
$B\to V\,l^+ l^-$ (where $V=K^*,\rho,\dots$ is a vector meson)
\cite{Beneke:2001at,Bosch:2001gv}. The resulting factorization formula 
is similar (but simpler) to that for $B$ decays into two light mesons. 
Therefore, the study of exclusive radiative transitions not only extends 
the range of applicability of the method, but also provides a new 
testing ground for the factorization idea.

Interestingly, the analysis of isospin-breaking effects in radiative $B$ 
decays gives a direct probe of power corrections to the factorization 
formula. Experimentally, it is found that 
\cite{Coan:2000kh,Ushiroda:2001sb,BaBar}
\[
   \Delta_{0-}\equiv
   \frac{\Gamma(\bar B^0\to\bar K^{*0}\gamma)
         -\Gamma(B^-\to\bar K^{*-}\gamma)}
        {\Gamma(\bar B^0\to\bar K^{*0}\gamma)
         +\Gamma(B^-\to\bar K^{*-}\gamma)} = 0.11\pm 0.07 \,,
\]
indicating (albeit with a large error) that isospin-breaking effects
could be as large as 10\% at the level of the decay amplitudes. Such 
effects are absent in the heavy-quark limit. A detailed theoretical
analysis of the leading power-suppressed contributions leads to the 
prediction $\Delta_{0-}=(8.0_{\,-\,3.2}^{\,+\,2.1})\%
\times(0.3/T_1^{B\to K^*})$ \cite{newpaper}, where $T_1^{B\to K^*}$ is a 
tensor form factor, whose value is expected to be close to 0.3. By far 
the largest contribution to the result comes from an annihilation 
contribution involving the $(V-A)\otimes(V+A)$ penguin operator $O_6$ in 
the effective weak Hamiltonian. Therefore, the quantity $\Delta_{0-}$ is 
a sensitive probe of the magnitude and sign of the ratio 
$C_6/C_{7\gamma}$ of Wilson coefficients.

The above discussion shows that in the Standard Model one indeed expects 
a sizeable isospin breaking in the $B\to K^*\gamma$ decay amplitudes, in 
agreement with the current central experimental value. If the agreement 
persists as the data become more precise, this would not only test the 
penguin sector of the effective weak Hamiltonian, but also provide a 
quantitative test of factorization at the level of power corrections.

\subsection{Tests of Factorization in $B\to\pi K,\pi\pi\,$ Decays} 

The factorization formula for $B$ decays into two light mesons is more
complicated because of the presence of the two types of contributions
shown in the graphs on the right-hand side in Figure~\ref{fig:fact}. The 
finding that these two topologies contribute at the same power in 
$\Lambda_{\rm QCD}/m_b$ is nontrivial \cite{Beneke:2001ev} and relies 
on the heavy-quark scaling law $F^{B\to L}(0)\sim m_b^{-3/2}$ for 
heavy-to-light form factors. Whereas this scaling law has been obtained
from several independent studies (see, e.g., 
\cite{Chernyak:1990ag,Ali:1994vd,Bagan:1998bp}), it is not as rigorously 
established as the corresponding scaling law for heavy-to-heavy form
factors. In the QCD factorization approach the kernels $T_{ij}^I(\mu)$
are of order unity, whereas the kernels $T_{ij}^{II}(\mu)$ contribute 
first at order $\alpha_s$. Numerically, the latter ones give corrections 
of about 10--20\% with respect to the leading terms. This is consistent 
with being of the same power but down by a factor of $\alpha_s$. 
Therefore, the scaling laws that form the basis of the QCD factorization
formula appear to work well empirically.

The factorization formula for $B$ decays into two light mesons can be 
tested best by using decays that have negligible amplitude interference. 
In that way any sensitivity to the value of the weak phase $\gamma$ is 
avoided. For a complete theoretical control over charmless hadronic 
decays one must control the magnitude of the tree topologies, the 
magnitude of the penguin topologies, and the relative strong-interaction 
phases between trees and penguins. It is important that these three key 
features can be tested separately. Once these tests are conclusive (and 
assuming they are successful), factorization can be used to constrain 
the parameters of the unitarity triangle. (Of course, alternative 
schemes such as pQCD \cite{Keum:2001ph} and ``charming penguins'' 
\cite{Ciuchini:1997hb} must face the same tests.)

\paragraph{Magnitude of the Tree Amplitude}
The magnitude of the leading $B\to\pi\pi$ tree amplitude can be probed in
the decays $B^\pm\to\pi^\pm\pi^0$, which to an excellent approximation 
do not receive any penguin contributions. The QCD factorization approach 
makes an absolute prediction for the corresponding branching ratio
\cite{Beneke:2001ev},
\[
   \mbox{Br}(B^\pm\to\pi^\pm\pi^0)
   = \Big[ 5.3_{\,-0.4}^{\,+0.8}\,(\mbox{pars.})
   \pm 0.3\,(\mbox{power}) \Big] \cdot 10^{-6}\times 
   \left[ \frac{|V_{ub}|}{0.0035}\,\frac{F_0^{B\to\pi}(0)}{0.28}
   \right]^2 ,
\]
which compares well with the experimental result
$(5.6\pm 1.5)\times 10^{-6}$ (see Table~7 in \cite{Beneke:2001ev} for a 
compilation of the experimental data on charmless hadronic $B$ decays). 
The theoretical uncertainties quoted are due to input parameter 
variations and to the modeling of the leading power corrections. An 
additional large uncertainty comes from the present error on $|V_{ub}|$ 
and the semileptonic $B\to\pi$ form factor. The sensitivity to these 
quantities can be eliminated by taking the ratio 
\[
   \frac{\Gamma(B^\pm\to\pi^\pm\pi^0)}
        {d\Gamma(\bar B^0\to\pi^+ l^-\bar\nu)/dq^2|_{q^2=0}}
   = 3\pi^2 f_\pi^2 \hspace{-0.6cm}
   \underbrace{|a_1^{(\pi\pi)}+a_2^{(\pi\pi)}|^2
               }_{1.33_{\,-0.11}^{\,+0.20}\,(\mbox{pars.})\pm 0.07\,
               (\mbox{power})} \hspace{-0.6cm}
   = (0.68_{\,-0.06}^{\,+0.11})\,\mbox{GeV}^2 \,.
\]
This prediction includes a sizeable ($\sim 25\%$) contribution of the 
hard-scattering term in the factorization formula (the lower graph on 
the right-hand side in Figure~\ref{fig:fact}). Unfortunately, this ratio 
has not yet been measured experimentally.

\paragraph{Magnitude of the $T/P$ Ratio}
The magnitude of the leading $B\to\pi K$ penguin amplitude can be probed 
in the decays $B^\pm\to\pi^\pm K^0$, which to an excellent approximation 
do not receive any tree contributions. Combining it with the measurement
of the tree amplitude just described, a tree-to-penguin ratio can be 
determined via the relation
\[
   \varepsilon_{\rm exp} = \left| \frac{T}{P} \right|
   = \mbox{tan}\theta_C\,\frac{f_K}{f_\pi}\, \left[ 
   \frac{2\mbox{Br}(B^\pm\to\pi^\pm\pi^0)}
        {\mbox{Br}(B^\pm\to\pi^\pm K^0)} \right]^{\frac12} 
   = 0.223\pm 0.034 \,.
\]
The quoted experimental value of this ratio is in good agreement with 
the theoretical prediction $\varepsilon_{\rm th}=0.24\pm 0.04\,
(\mbox{pars.})\pm 0.04\,(\mbox{power})\pm 0.05\,(V_{ub})$ 
\cite{Beneke:2001ev}, which is independent of form factors but 
proportional to $|V_{ub}/V_{cb}|$. This is a highly nontrivial test of 
the QCD factorization approach. Recall that when the first measurements 
of charmless hadronic decays appeared several authors remarked that the 
penguin amplitudes were much larger than expected based on naive 
factorization models. We now see that QCD factorization reproduces 
naturally (i.e., for central values of all input parameters) the correct 
magnitude of the tree-to-penguin ratio. This observation also shows that 
there is no need to supplement the QCD factorization predictions in an 
ad hoc way by adding enhanced phenomenological penguin amplitudes, such 
as the ``nonperturbative charming penguins'' introduced in 
\cite{Ciuchini:1997hb}. In their most recent paper 
\cite{Ciuchini:2001pk}, the advocates of charming penguins parameterize 
the effects of these animals in terms of a nonperturbative ``bag 
parameter'' $\hat B_1=(0.13\pm 0.02)\,e^{i(188\pm 82)^\circ}$ fitted to 
the data on charmless decays. By definition, this parameter contains the 
contribution from the perturbative charm loop, which is calculable in 
QCD factorization. Using the factorization approach as described in 
\cite{Beneke:2001ev} we find that $\hat B_1^{\rm fact}
=(0.09_{\,-0.02-0.02}^{\,+0.03+0.04})\,e^{i(185\pm 3\pm 21)^\circ}$, 
where the errors are due to input parameter variations and the estimate 
of power corrections. The perturbative contribution to the central value 
is 0.08; the remaining 0.01 is mainly due to weak annihilation. We 
conclude that, within errors, QCD factorization can account for the 
``charming penguin bag parameter'', which is, in fact, dominated by 
short-distance physics.

\begin{table}
\label{tab:ACPs}
\vspace{0.3truecm}
\caption{Direct CP asymmetries in $B\to\pi K$ decays}
\begin{tabular}{c|c|ccc}
\hline
 & Experiment & \multicolumn{3}{c}{Theory} \\
 & \cite{Chen:2000hv,Abe:2001hs,Aubert:2001hs,Aubert:2001qj}
 & Beneke et al.\ \cite{Beneke:2001ev}
 & Keum et al.\ \cite{Keum:2001ph}
 & Ciuchini et al.\ \cite{Ciuchini:1997hb}
 \\
\hline
$A_{\rm CP}(\pi^+ K^-)$~(\%) & $-4.8\pm 6.8$ & $5\pm 9$ & $-18$
 & $\pm(17\pm 6)$ \\
$A_{\rm CP}(\pi^0 K^-)$~(\%) & $-9.6\pm 11.9$ & $7\pm 9$ & $-15$
 & $\pm(18\pm 6)$ \\
$A_{\rm CP}(\pi^-\bar K^0)$~(\%) & $-4.7\pm 13.9$ & $1\pm 1$ & $-2$
 & $\pm(3\pm 3)$ \\
\hline
\end{tabular}
\end{table}

\paragraph{Strong Phase of the $T/P$ Ratio}
QCD factorization predicts that (most) strong-interaction phases in 
charmless hadronic $B$ decays are parametrically suppressed in the 
heavy-quark limit, i.e., 
$\sin\phi_{\rm st}=O[\alpha_s(m_b),\Lambda_{\rm QCD}/m_b]$. This implies
small direct CP asymmetries since, e.g., $A_{\rm CP}(\pi^+ K^-)\simeq 
-2\,|\frac{T}{P}|\sin\gamma\,\sin\phi_{\rm st}$. The suppression results 
as a consequence of systematic cancellations of soft contributions, 
which are missed in phenomenological models of final-state interactions. 
In many other schemes the strong-interaction phases are predicted to be 
much larger, and therefore larger CP asymmetries are expected. 
Table~\ref{tab:ACPs} shows that first experimental data provide no 
evidence for large direct CP asymmetries in $B\to\pi K$ decays. However, 
the errors are still too large to draw a definitive conclusion that 
would allow us to distinguish between different theoretical predictions.

\subsection{Remarks on Sudakov Logarithms}

In recent years, Li and collaborators have proposed an alternative
scheme for calculating nonleptonic $B$ decay amplitudes based on a 
perturbative hard-scattering approach \cite{Keum:2001ph}. From a 
conceptual point of view, the main difference between QCD factorization 
and this so-called pQCD approach lies in the latter's assumption that 
Sudakov form factors effectively suppress soft-gluon exchange in 
diagrams such as those shown in the graphs on the right-hand side in 
Figure~\ref{fig:fact}. As a result, the $B\to\pi$ and $B\to K$ form
factors are assumed to be perturbatively calculable. This changes the 
counting of powers of $\alpha_s$. In particular, the nonfactorizable 
gluon exchange diagrams included in the QCD factorization approach, 
which are crucial in order to cancel the scale and scheme-dependence in
the predictions for the decay amplitudes, are formally of order 
$\alpha_s^2$ in the pQCD scheme and consequently are left out. Thus, to
the considered order there are no loop graphs that could give rise to 
strong-interaction phases in that scheme. (However, in \cite{Keum:2001ph}
large phases are claimed to arise from on-shell poles of massless 
propagators in tree diagrams. These phases are entirely dominated by 
soft physics. Hence, the prediction of large direct CP asymmetries in 
the pQCD approach rests on assumptions that are strongly model 
dependent.)

The assumption of Sudakov suppression in hadronic $B$ decays is 
questionable, because the relevant ``large'' scale 
$Q^2\sim m_b\Lambda_{\rm QCD}\sim 1$\,GeV$^2$ is in fact not large 
for realistic $b$-quark masses. Indeed, one finds that the pQCD 
calculations are very sensitive to details of the $p_\perp$ dependence 
of the wave functions \cite{Chris}. This sensitivity to infrared physics 
invalidates the original assumption of an effective suppression of soft 
contributions. The argument just presented leaves open the conceptual 
question whether Sudakov logarithms are relevant in the asymptotic limit 
$m_b\to\infty$. This question has not yet been answered in a 
satisfactory way.

\section{New Constraints on the Unitarity Triangle}

The QCD factorization approach, combined with a conservative estimate of 
power corrections, offers several new strategies to derive constraints 
on CKM parameters. This has been discussed at length in 
\cite{Beneke:2001ev}, to which we refer the reader for details. Some of 
these strategies will be illustrated below. Note that the applications 
of QCD factorization are not limited to computing branching ratios. The
approach is also useful in combination with other ideas based on flavor 
symmetries and amplitude relations. In this way, strategies can be found
for which the residual hadronic uncertainties are simultaneously 
suppressed by three small parameters, since they vanish in the 
heavy-quark limit ($\sim\Lambda_{\rm QCD}/m_b$), the limit of SU(3) 
flavor symmetry ($\sim (m_s-m_q)/\Lambda_{\rm QCD}$), and the large-$N_c$
limit ($\sim1/N_c$).

\begin{figure}
\label{fig:NR}
\includegraphics[height=.3\textheight]{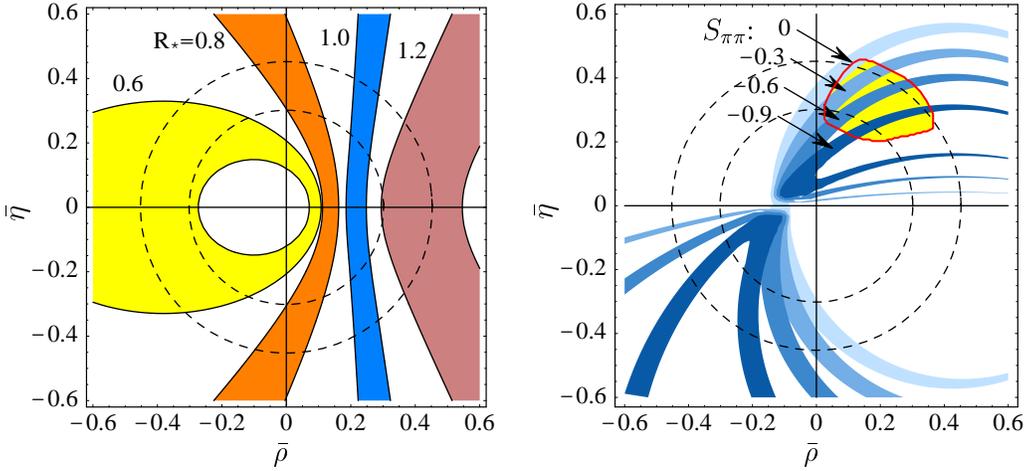}
\caption{Left: Allowed regions in the $(\bar\rho,\bar\eta)$ plane 
corresponding to $\varepsilon_{\rm exp}=0.22$ and different values of 
the ratio $R_*$ as indicated. The widths of the bands reflect the total 
theoretical uncertainty. The current experimental values are 
$\varepsilon_{\rm exp}=0.22\pm 0.03$ and $R_*=0.71\pm 0.14$. Right: 
Allowed regions in the $(\bar\rho,\bar\eta)$ plane corresponding to 
different values of the mixing-induced CP asymmetry $S_{\pi\pi}$. The 
widths of the bands reflect the total theoretical uncertainty. The 
corresponding bands for positive values of $S_{\pi\pi}$ are obtained by 
a reflection about the $\bar\rho$ axis. The bounded light area is the 
allowed region obtained from the standard global fit of the unitarity 
triangle \protect\cite{Hocker:2001xe}.}
\end{figure}

\paragraph{Determination of $\gamma$ with minimal theory input}
Some years ago, Rosner and the present author have derived a bound on 
$\gamma$ by combining measurements of the ratios 
$\varepsilon_{\rm exp}=|T/P|$ and 
$R_*=\frac12\,\Gamma(B^\pm\to\pi^\pm K^0)/\Gamma(B^\pm\to\pi^0 K^\pm)$ 
with the fact that for an arbitrary strong phase
$-1\le\cos\phi_{\rm st}\le 1$ \cite{Neubert:1998pt}. The 
model-independent observation that $\cos\phi_{\rm st}=1$ up to 
second-order corrections to the heavy-quark limit can be used to turn 
this bound into a determination of $\gamma$ (once $|V_{ub}|$ is known). 
The resulting constraints in the $(\bar\rho,\bar\eta)$ plane, obtained
under the conservative assumption that $\cos\phi_{\rm st}>0.8$ 
(corresponding to $|\phi_{\rm st}|<37^\circ$) are shown in the 
left-hand plot in Figure~\ref{fig:NR} for several illustrative values
of the ratio $R_*$. Note that for $0.8<R_*<1.1$ (the range preferred 
by the Standard Model) the theoretical uncertainty reflected by the
widths of the bands is smaller than for any other constraint on 
$(\bar\rho,\bar\eta)$ except for the one derived from the $\sin2\beta$
measurement. With present data the Standard Model is still in good 
shape, but it will be interesting to see what happens when the 
experimental errors are reduced.

\paragraph{Determination of $\,\sin 2\alpha$}
With the help of QCD factorization it is possible to control the 
``penguin pollution'' in the time-dependent CP asymmetry in 
$B\to\pi^+\pi^-$ decays, defined such that 
$S_{\pi\pi}=\sin2\alpha\cdot[1+O(P/T)]$. This is illustrated in the
right-hand plot in Figure~\ref{fig:NR}, which shows the constraints 
imposed by a measurement of $S_{\pi\pi}$ in the $(\bar\rho,\bar\eta)$ 
plane. It follows that even a result for $S_{\pi\pi}$ with large 
experimental errors would imply a useful constraint on the unitarity 
triangle. A first, preliminary measurement of the asymmetry has been 
presented by the BaBar Collaboration this summer. Their result is 
$S_{\pi\pi}=0.03_{\,-0.56}^{\,+0.53}\pm 0.11$ \cite{Aubert:2001qj}.

\begin{figure}
\label{fig:UTfit}
\includegraphics[width=.58\textwidth]{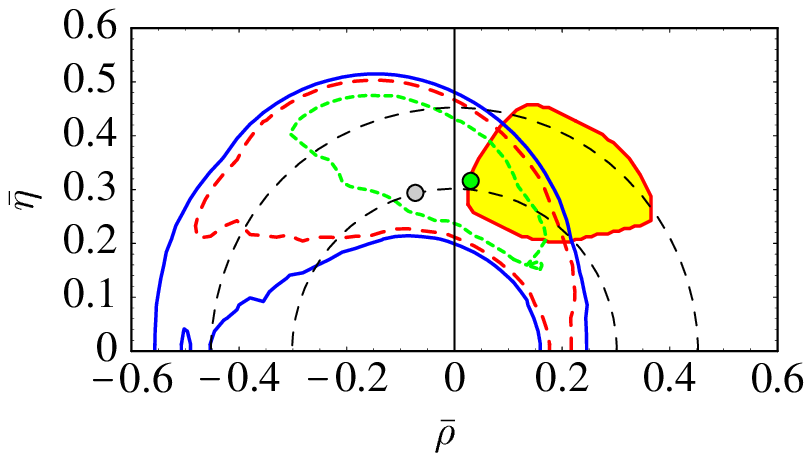}
\includegraphics[width=.4\textwidth]{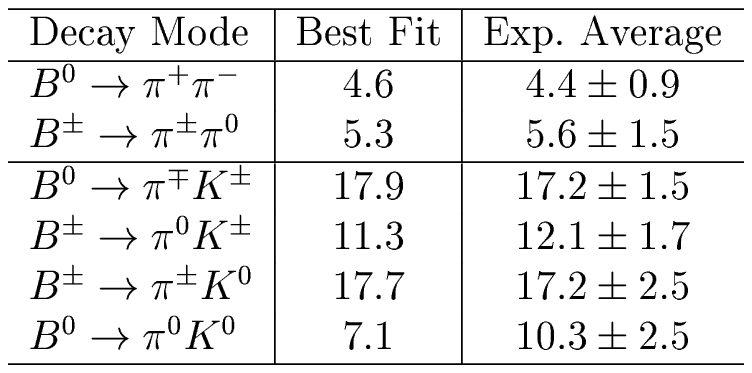}
\caption{95\% (solid), 90\% (dashed) and 68\% (short-dashed) confidence 
level contours in the $(\bar\rho,\bar\eta)$ plane obtained from a global 
fit of QCD factorization results to the CP-averaged $B\to\pi K,\pi\pi$ 
branching fractions. The dark dot shows the overall best fit, whereas 
the light dot indicates the best fit for the default choice of all 
theory input parameters. The table compares the best fit values for the 
various CP-averaged branching fractions (in units of $10^{-6}$) with the 
world average data.}
\end{figure}

\paragraph{Global Fit to $B\to\pi K,\pi\pi$ Branching Ratios}
Various ratios of CP-averaged $B\to\pi K,\pi\pi$ branching fractions 
exhibit a strong dependence on $\gamma$ and $|V_{ub}|$, or equivalently,
on the parameters $\bar\rho$ and $\bar\eta$ of the unitarity triangle. 
From a global analysis of the experimental data in the context of the
QCD factorization approach it is possible to derive constraints in the 
$(\bar\rho,\bar\eta)$ plane in the form of regions allowed at various
confidence levels. The results are shown in Figure~\ref{fig:UTfit}. The 
best fit of the QCD factorization theory to the data yields an excellent 
$\chi^2/n_{\rm dof}$ of less than 0.5. 
(We should add at this point that we disagree with the implementation of
our approach presented in \cite{Ciuchini:2001pk} and, in particular, 
with the numerical results labeled ``BBNS'' in Table~II of that paper,
which led the authors to the premature conclusion that the ``theory of
QCD factorization ... is insufficient to fit the data''. Even 
restricting $(\bar\rho,\bar\eta)$ to lie within the narrow ranges
adopted by these authors, we can find parameter sets for which QCD 
factorization fits the data with a good $\chi^2/n_{\rm dof}$ of less 
than 1.5.) 

The results of this global fit are compatible with the standard CKM fit 
using semileptonic decays, $K$--$\bar K$ mixing and $B$--$\bar B$ mixing
($|V_{ub}|$, $|V_{cb}|$, $\epsilon_K$, $\Delta m_d$, $\Delta m_s$, 
$\sin 2\beta$), although the fit prefers a slightly larger value of 
$\gamma$ and/or a smaller value of $|V_{ub}|$. The combination of the
results from rare hadronic $B$ decays with $|V_{ub}|$ from semileptonic 
decays excludes $\bar\eta=0$ at 95\% CL, thus showing first evidence for 
the existence of a CP-violating phase in the bottom sector. In the near 
future, when the data become more precise, this will provide a powerful 
test of the CKM paradigm.

\section{Outlook}

The QCD factorization approach provides the theoretical framework for a 
systematic analysis of hadronic and radiative exclusive $B$ decay 
amplitudes based on the heavy-quark expansion. This theory has already 
passed successfully several nontrivial tests, and will be tested more 
thoroughly with more precise data. A new effective field-theory language 
appropriate to QCD factorization is emerging in the form of the 
collinear--soft effective theory. Ultimately, the developments reviewed
in this talk may lead to theoretical control over a vast variety of 
exclusive $B$ decays, giving us new constraints on the unitarity 
triangle.

\begin{theacknowledgments}
I wish to thank the organizers of {\em QCD@Work\/} and {\em Heavy 
Flavours 9\/} for the invitation to deliver this talk, and for 
arranging these two conferences in a style and setting which encouraged 
many lively discussions. This work was supported in part by the National 
Science Foundation.

The second version of this talk was delivered on 11 September 2001, a 
few hours after terrorist attacks hit various targets in the U.S. Our 
thoughts and sympathy are with the many innocent victims of this 
tragedy.
\end{theacknowledgments}

\doingARLO[\bibliographystyle{aipproc}]
 {\ifthenelse{\equal{\AIPcitestyleselect}{num}}
 {\bibliographystyle{arlonum}}
 {\bibliographystyle{arlobib}}
}
\bibliography{mytalk}

\end{document}